\documentclass[conference]{IEEEtran}

\ifCLASSOPTIONcompsoc
\else
\fi
\ifCLASSINFOpdf
  \usepackage[pdftex]{graphicx}
\else
  \usepackage[dvips]{graphicx}
\fi
\usepackage[cmex10]{amsmath}
\ifCLASSOPTIONcompsoc
\usepackage[tight,normalsize,sf,SF]{subfigure}
\else
\usepackage[tight,footnotesize]{subfigure}
\fi

\usepackage{enumitem}
\usepackage{hyperref}
\usepackage{xcolor,listings,courier}

\lstnewenvironment{Code}[1][]
 {\lstset{language=[LaTeX]TeX}\lstset{%
      breaklines=true,
      framesep=5pt,
      basicstyle=\scriptsize\ttfamily,
      showstringspaces=false,
      keywordstyle=\footnotesize\textcolor{blue},
      stringstyle=\color{orange},
     commentstyle=\color{black},
     rulecolor=\color{gray!10},
      breakatwhitespace=true,
     showspaces=false,  
     xleftmargin=0pt,
     xrightmargin=5pt,
     aboveskip=0pt, 
     belowskip=0pt,  
      backgroundcolor=\color{gray!15}, #1,
}}
{}

\begin{document}

\title{Multi-tenant Pub/Sub Processing\\for Real-time Data Streams}
%
%
%
%

\author{
\IEEEauthorblockN{\'{A}lvaro~Villalba, David~Carrera~\IEEEmembership{Member,~IEEE}}
\IEEEauthorblockA{Technical University of Catalonia (UPC) \\
Barcelona Supercomputing Center (BSC)}\\
\IEEEauthorblockA{Email: \{alvaro.villalba, david.carrera\}@bsc.es }
}

\IEEEcompsoctitleabstractindextext{%
\begin{abstract}
Devices and sensors generate streams of data across a diversity of locations
and protocols. That data usually reaches a central platform that is used to
store and process the streams. Processing can be done in real time, with
transformations and enrichment happening on-the-fly, but it can also happen
after data is stored and organized in repositories. In the former case, stream
processing technologies are required to operate on the data; in the latter
batch analytics and queries are of common use.

This paper introduces a runtime to dynamically
construct data stream processing topologies based on user-supplied code.
These dynamic topologies are built on-the-fly using a data subscription model
defined by the applications that consume data. Each user-defined processing unit
is called a Service Object. Every Service Object consumes input data streams
and may produce output streams that others can consume.
The subscription-based programing model enables multiple users to deploy their own data-processing
services. The runtime does the dynamic forwarding of data and execution of
Service Objects from different users.
Data streams can originate in real-world devices or they can be the outputs of
Service Objects.

The runtime leverages Apache STORM
for parallel data processing, that combined with dynamic user-code
injection provides multi-tenant stream processing topologies. 
In this work we describe the runtime, its features and implementation
details, as well as we include a performance evaluation of some of its core
components.
\end{abstract}

\begin{IEEEkeywords}
Big Data, Analytics, Stream Processing, Real-time Data Processing, Programming Models, Internet of Things, IoT 
\end{IEEEkeywords}
}

\maketitle

\IEEEdisplaynotcompsoctitleabstractindextext

%
\IEEEpeerreviewmaketitle

\section{Introduction}

In the last years, Big Data and Internet of Things (IoT) platforms are clearly
converging in terms of technologies, problems and approaches. IoT ecosystems
generate a vast amount of data that needs to be stored and processed, becoming a
Big Data problem. Devices and sensors generate streams of data across a
diversity of locations and protocols that in the end reach a central platform
that is used to store and process it. Processing can be done in real
time, with transformations and enrichment happening on-the-fly, but it can also
happen after data is stored and organized in repositories. 

This situation implies an increasing demand for
advanced data streams management and processing platforms. Such platforms require
multiple protocols support for extended connectivity with the
objects. But also need to exhibit uniform internal data organization and
advanced data processing capabilities to fulfill the demands of the application
and services that consume these streams of data.

To provide answer to this growing demand, ServIoTicy\footnote{servioticy.com} is
a state-of-the-art platform for hosting real-time data stream workloads in
the Cloud. It provides multi-tenant data stream processing capabilities, a REST
API, data analytics, advanced queries and multi-protocol support in a
combination of advanced data-centric services. 
The main focus of ServIoTicy is to provide a rich set of features to store and
process data through its REST API, allowing objects, services and humans to
access the information produced by the devices connected to the platform.
ServIoTicy allows for a real time processing of device-generated data, and
enables for simple creation of data transformation pipelines using user
generated logic. Unlike traditional service composition approaches, usually
focused on addressing the problems of functional composition of existing
services, one of the goals of the ServIoTicy is to focus on data processing
scalability.
Other components that can be connected to ServIoTicy provide added capabilities
to automatically create compositions of high-level services using existing
tools~\cite{iserve}. 

The core of the ServIoTicy runtime relies on a novel programing model
that allows users to dynamically
construct data stream processing topologies based on user-supplied code.
These topologies are built on-the-fly according to a data subscription model
defined by the applications that consume data. Once a stream subscriber finishes
its work, it is freed from the platform until it is needed again. Each user-defined processing unit
is called a Service Object (SO). Every Service Object consumes input data streams
and may produce output streams that others can consume. Data streams can
originate in real-world devices or they can be outputs of Service Objects
deployed in the platform.

Advanced streaming and analytics platforms such as ServIoTicy are complex pieces
of software that integrate a large set of components under the hood. They hide
their complexity behind simple REST APIs and multi-protocol channels, but the
reality is that their deployment and configuration is complex. ServIoTicy
leverages Apache STORM runtime for parallel data processing, that combined with
dynamic user-code injection provides multi-tenant stream processing topologies. 

This paper provides insights on the performance properties of ServIoTicy as an
starting point for the construction of advanced cloud provisioning strategies
and algorithms. The work presented here focuses on the processing topologies
built in ServIoTicy, although some details about other platform components are
also provided.

Security is one of the main concerns on IoT platforms because they deal with big
amounts of sensitive data. Although the applied security policies are not in
the scope of this paper, there has been efforts in that matter. Each update contains
provenance data including the data owners and the operations that has been applied. 
The provenance data is used with a security policy manager to decide if an application 
can make use of the update.


The source code of ServIoTicy is freely available as an open source
project\footnote{https://github.com/servioticy} in GitHub.
The platform is also available for single node testing as a vagrant box,
downloadable from a github
repository\footnote{https://github.com/servioticy/servioticy-vagrant}.

The main contributions of this paper are:
\begin{itemize}
	\item A technique for user-code injection on a data stream
	processing runtime that allows for multi-tenant stream
	processing on-the-fly. This runtime is the core of the ServIoTicy platform.
	\item An insight on the performance of the code-injection technique, including
	response time end-to-end in a processing pipeline and across stages.
\end{itemize}

The next sections of the paper are organized as follows:
Section~\ref{sec:architecture} introduces the general architecture and
components of the platform; 
Section~\ref{sec:abstractions} introduces a set of abstractions defined in
ServIoTicy for managing data associated to objects;
Section~\ref{sec:strprocessing} describes in detail the stream processing
runtime of ServIoTicy;
Section~\ref{sec:experiments} presents the evaluation methodology and the
experiment included in the paper; Finally, Section~\ref{sec:related} goes
through the related work and Section~\ref{sec:futurework} provides some
conclusions and future lines of work.

\section{Architecture of ServIoTicy}
\label{sec:architecture}

The Front-End of platform is a Web Tier that
implements the REST API that sits at the core of ServIoTicy. The API
contains parts of the logic of the Service Objects and Data Processing Pipelines,
related to authentication, data storage and data retrieval actions. The Stream
Processing Topology is responsible for the execution of the code associated to
Data Processing pipes as well as the forwarding of data across Service Objects
and to external entities (e.g. external subscribers that want data forwarded on
real-time using a push model on top of MQTT or STOMP). Finally, the data
Back-End includes the Data Store that provides scalable, distributed and
fault-tolerant properties to ServIoTicy, and the Indexing Engine that provides
search capabilities across sensors data using different criteria, like
timestamps, string patterns or geo-location. In this section we describe in more
detail the main properties of each component of the ServIoTicy architecture.


\subsubsection{Web Tier}
The Web Tier for the REST API is composed of a Servlets Container and a REST
Engine. As a HTTP Web Server and Java Servlet container we use
Jetty~\cite{jetty}. Jetty is often used for machine-to-machine communications, usually
within larger software frameworks. 
As a JSON processor we use Jackson~\cite{jackson}, which is a high-performance suite of data-processing
tools for Java, including the flagship JSON parsing and generation library, as well as additional modules. The Jackson
Project also has handlers to add data format support for JAX-RS implementations
like Jersey.

\subsubsection{Stream Processing Topology}

The Stream Processing Topology is implemented on top of Apache
STORM~\cite{storm}, which is a state-of-the-art stream processing runtime.
Out-of-the-box, STORM provides the availability to build topologies composed of
spouts (sources of data) and bolts (processing units). Topologies are static
after their deployment, and data keeps flowing through their bolts until
the topology is stopped. STORM provides auto-scaling capabilities that make it
particularly suitable for cloud deployments.
Note that in case that a different topology is needed, the user needs
to stop the running topology and deploy the new one. This situation will not 
affect the final platform, as it will be explained in more detail in following
sections.
 The Stream Processing Topology also requires the support of a queuing
system that will act as the spout for the STORM topology. In ServIoTicy, this is
implemented using Kestrel~\cite{kestrel}.

\subsubsection{Data Store}

A distributed data store is used to keep track of all the object produced 
data. For that purpose, CouchBase~\cite{couchbase} has been 
chosen as the data store because it provides the benefits of NoSQL data stores 
(highly distributed, high-availability properties, scalable), and it is document 
oriented (which fits well for many different data sources and formats). 
Couchbase has native support for JSON documents. The definition of all Service
Objects in ServIoTicy and their associated streams are stored as JSON documents
in Couchbase.

\subsubsection{Data Indexing}

The search infrastructure to resolve
queries is provided by an underlying component that performs
high-performance indexing and search operations. In particular
Elasticsearch~\cite{elasticsearch} is leveraged as it is one of the most
powerful and extended search engines that can be integrated with scalable data
back-ends (in particular Couchbase).
The integration between Couchbase and Elasticsearch enables full-text search,
indexing and querying and real-time analytics for variety of use cases such as a
content store or aggregation of data from different data sources.

\subsubsection{Multi-Protocol Brokerage}
In an attempt to make ServIoTicy platform more accessible to udevices,
particularly those with less computing capacity or with more power constraints,
the REST API is also reachable using other protocols and transports. In
particular, STOMP over TCP and WebSockets, and
MQTT over TCP are also available.
All these features are implemented in ServIoTicy using a
combination of newly developed bridges between components and Apache
Apollo~\cite{apollo} as the core message brokering engine.

\section{Abstractions used in ServIoTicy}
\label{sec:abstractions}

Several abstractions are used in ServIoTicy to embrace the different entities
involved in the existence of IoT ecosystems.

\begin{itemize}[leftmargin=*]

  \item Web Object: The platform gathers information from objects, either connected to the
  Internet or not. The group of objects not directly connected to the Internet
  (e.g. a bottle of wine with a RFID or NFC tag) will need a proxy to represent
  them in the ServIoTicy. There is also a group of objects which may have
  network capabilities, but limited programmability and support for advanced
  network protocols. These devices, such as simple sensors, still will need the
  use of proxies to be able to communicate with ServIoTicy. Finally, there is a
  group of advanced devices (so-called Smart Objects, such as a Smart Phone,
  tablet, or an Arduino device) that already hold the capabilities to talk to
  the COMPOSE platform directly. Each one of the mentioned objects (enabled
  with a communications proxy when needed) is known as a Web Object (WO) in
  ServIoTicy. Web Objects are physical objects sitting on the edge of the
  ServIoTicy and capable of keeping for example HTTP-based bi-directional
  communications, such that the object will be able to both send data to the
  platform and receive activation requests and notifications.
  Not all such objects will support the same set of operations, but a minimum
  subset will have to be guaranteed to make them usable to ServIoTicy.
    
  \item Service Object: Service Objects are standard internal ServIoTicy
  representations of Web Objects. ServIoTicy specifies an API 
  by which it expects to communicate with the
  Web Objects, in order to obtain data from them, or set data within them. That
  API can be embedded directly in the Objects or can be provided by a mediating
  proxy that will connect the Objects to their corresponding ServIoTicy Service
  Objects. This entity serves mainly for data management purposes and has a
  well-defined and closed API. That API is needed in order to streamline and
  standardize internal access to Service Objects, which can in turn represent a
  variety of very different Web Objects providing very different capabilities.
  ServIoTicy, in an effort to embrace as many IoT transports as possible, allows
  Web Objects to interact with their representatives in the Platform (the
  Service Objects) using a set of well-known protocols: HTTP, STOMP~\cite{stomp}
  over TCP, STOMP over WebSockets~\cite{websockets}, and MQTT~\cite{mqtt}
  over TCP.
  
  \item Data Processing Pipeline: A Data Processing Pipeline is a data service and
  aggregation mechanism, which relies on the data processing and management
  back-end component to provide complex computations resulting from
  subscriptions to different Service Objects as data sources. This construct can
  support pseudo-real time data stream transformations, combined with queries
  concerning historical data. Data analytics code defined by the user may be
  provided as well. The end result of a Data Processing Pipeline is inserted into
  the ServIoTicy registry along with its description and may be used by higher
  level constructs as yet another kind of Service Object building block. Just
  like a Service Object, this entity serves mainly for data management purposes
  and has a well-defined and closed API.
  
  \item Subscription: Data subscriptions are a mechanism in ServIoTicy that
  allow Service Objects, Data Processing Pipelines and external data consumers to
  get data updates automatically and asynchronously forwarded for further
  processing.
  
  \item Sensor Update: Sensor Updates are the unit of data sent by a Web Object
  to its Service Object. It contains the different synchronously sensed values
  and a timestamp that is maintained all over the pipelines. A subscription or a 
  query to a Service Object will get the 
  data in this format.
  
\end{itemize}

\section{Data Processing Pipelines}
\label{sec:strprocessing}
Service Objects store their associated data in abstractions called
\emph{streams}. The unit of data that can be observed for one stream
is called a \emph{Sensor Update} (SU). Applications can subscribe to or query
data associated to any stream.
Streams can be of two different types:
\begin{itemize}[leftmargin=*]
	\item Simple data streams store data generated in the physical world by a sensing
	device, assuming that a device with N sensors will generate N streams of data 
	that will be grouped in a Service Object abstraction that represents the
	device.
	\item Composite data streams represent transformations (aggregate, merge,
	filter or join, among other possibilities) performed on other data sources
	(either by devices located in the physical world or by Service Objects existing
	in the ServIoTicy platform). They can be thought about as a virtual
	(non-physical) sensor of the SO.
\end{itemize}


From an API perspective there is no difference between a simple stream and a
composite stream, as they both support queries and subscriptions.
Therefore, the inputs of composite stream can be streams or other composite
streams. These chained transformations of SUs are called \emph{Data Processing
Pipelines}.

Listing~\ref{text:composition} is a snippet from a SO descriptor that illustrates the case 
of a composite stream that takes temperature reads in Fahrenheit degrees as input SUs and produces 
temperatures in Celsius degrees as outputs if and only if the temperature
is below $0\,^{\circ}{\rm C}$. Note how the \emph{current-value} of the stream
is calculated first by transforming the $\,^{\circ}{\rm F}$ into $\,^{\circ}{\rm C}$, and afterwards a {\it
post-filter} is used to discard any outputs that would correspond to positive temperatures.
The following sections will describe in more detail the purpose of the elements
of this example and their semantics.

\vspace{2ex}
\begin{Code}[caption=Example of data transformation and filtering: convert from
$^{\circ}{\rm F}$ to $^{\circ}{\rm C}$ and retaining only non-freezing
temperatures\vspace{1.5ex}, label=text:composition][th]
"streams":{ 
 "frozencelsius": { 
  "channels": {
   "temp": {
     "type": "number",
     "current-value": "({$fahrenheit.channels.temp.
                          current-value} - 32) / 1.8",
     
     "post-filter":   "({$result.channels.temp.
                          current-value} < 0)" 
} } } }
\end{Code}

\subsection{Data Structures}
The structure of a Sensor Update that corresponds to a given stream is basically
composed of a series of \emph{Channels} associated to the dimensions of the data represented
by the stream (e.g. a geo-location stream may contain two channels representing
the latitude and the longitude correspondingly), and a timestamp reported by
the data source as the time at which the Sensor Update was generated. 

The composite stream structure is similar to the structure 
of a SU. It contains channels, and each channel contains a so-called
'current-value' field that represents the output value that the composite stream will
emit after ingesting a new SU, assuming that the output is not filtered. In a SO
document, the content of a 'current-value' field is a string with a JavaScript 
variable assignment using any mix of basic operator and functions from the Math object,
String object, Array object, as well as shorthand conditional expressions (a = b
? true : false). The result of the assignment to 'current-value' will always be
numeric, a Boolean, a string or an array of the previous types. It will be stored
and emitted to its subscribers.

\subsection{Stages of the Processing Pipeline}
\label{sec:stages}
Once a SU reaches a composite stream as one of its inputs, it goes through a number 
of stages in order to transform it into a new output SU. This process of ingesting 
a SU and processing it until a new SU is produced can be summarized as the
following set of stages:


\begin{enumerate}[leftmargin=*]
	\item Subscriber dispatching: A sensor update gets into the processing pipeline,
	along with its origin information. This stage looks for the subscribers of its
	origin and if they are composite streams, they are requested and sent to the next
	stage with the SU.
	
	\item Data Fetching: The composite stream may need access to the data stored by
	other streams that are inputs involved in the data transformation. In each
	stage, the sources needed by the stream are queried and their data made
	available for the rest of the stages, altogether with the original SU.
	References to fields on the Sensor Updates are made using JSONPaths.
	
%
%
	
	\item Transformation \& filtering:  Data transformation is performed by taking all the SUs
	extracted from all the data sources, and operating on their associated data
	using JavaScript algebraic operations and its Math object functions, String object
	operations, Array object operations, and boolean operations, to finally obtain a
	single value for the new SU to emit. Also, before and after the transformation
	SUs are discarded if a defined filter assertion
	is false, and no further stages would follow.
	
	\item Store, trigger actions and emit: Finally, the generated SU gets stored and
	emitted to the stream subscribers.
	Additionally, in this final stage, actions to be sent back to SOs are
	triggered. Such actions will end up being sensor actuations that will be driven
	through the WOs that embed the actual physical objects. 
\end{enumerate}

In ServIoTicy, basic physical object actuation is driven through SOs. When
a SO gets an action invoked through the SO actions API, the action is initiated on the
corresponding WO, that will act as a proxy for the physical actuator.
If a user needs to be able to manually request the execution of a composite
action (involving multiple SOs), it is necessary to create a SO that includes
the desired action and references to the individual SOs representing each of
the physical objects to be actuated, so that the composite action can be
properly triggered.

\subsection{Design Principles}

The data processing pipelines introduced in this work are intended to be scalable
in accordance with other works found in the literature~\cite{8requirements}.
In particular, the key design principles considered for the data processing pipelines were:

\begin{itemize}[leftmargin=*]
 \item Event-driven: A new SU calculation is triggered in a stream when it 
receives a SU.
 \item Lock-free: A stream that needs of several different SUs to generate a new 
one will not lock until all of them are received. It makes use of the received SU, 
and queries the last SUs from the other needed streams.
 \item Real-time data processing oriented: Each new SU is processed individually 
  without waiting for a batch.
\end{itemize}

The approach followed by ServIoTicy is an asynchronous model for which only one of 
the sources needs to issue a sensor update to trigger the processing of the 
composite stream. It enforces a high rate of updates and avoids locking the generation of 
new updates because one sensor is idle. This situation would lock an entire pipeline.



\begin{figure}[h]
\centering
\includegraphics[width=0.40\textwidth]{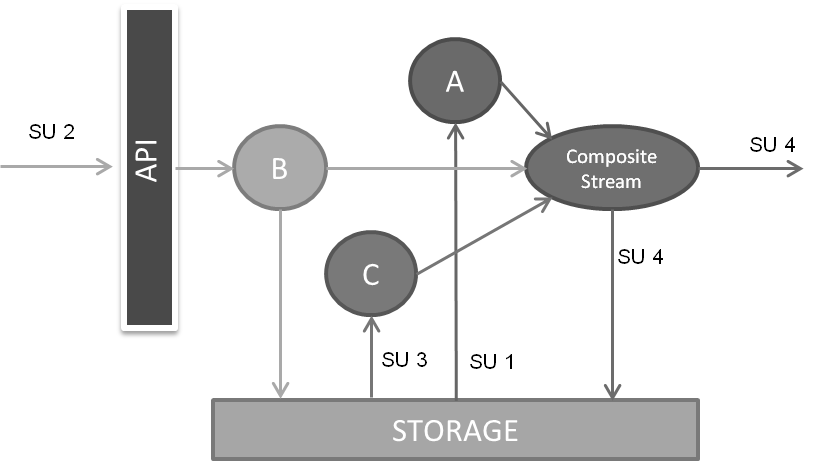}
\caption{Lock-free asynchronous model used in ServIoTicy}
\label{fig:workflow-cs-principles}
\end{figure}

Figure~\ref{fig:workflow-cs-principles}
illustrates the actual approach implemented in ServIoTicy using a lock-free scalable model. 
An update owned by stream \emph{B} is sent to ServIoTicy through the API and is stored.
A composite stream is subscribed to the streams \emph{A}, \emph{B} and \emph{C}, and so it 
receives their outputs SUs as inputs. It generates a new 
SU, stores it and becomes sent to further composite streams if any.
In this particular case, the generation of \emph{SU 4} also requires of \emph{SU 1} and \emph{SU 3}, so
the composite stream queries them to streams \emph{A} and \emph{C}. A single
event (receiving a SU) generates a single output in the composite stream.

\subsection{Time, Data Consistency and Efficiency}
A composite stream can take as inputs the most recent SU from any stream
declared in the platform, either from its own Service Object or from any other
Service Object. In the context of a particular data stream, that receives SUs as
inputs and stores data associated to its outputs in the platform, some restrictions
need to be in place to keep chronological consistency of the data being
produced by a given composite stream.

\vspace{2ex}
\begin{Code}[caption=Algorithm used to generate new updates,label=list:consistency][th]
def generateNewUpdate(receivedUpdate, currentStream, streamSubscriptions):

    previousSelfUpdateFuture = getLastUpdateAsync(currentStream)
    originStream = receivedUpdate.getStream(receivedUpdate)
    streamSubscriptions.remove(originStream)
    queriedUpdatesFuture = getLastUpdatesAsync(streamSubscriptions)
    
    // Block to receive the stream last update
    previousSelfUpdate = previousSelfUpdateFuture.get()
    if receivedUpdate.getTimestamp() <= previousSelfUpdate.getTimestamp()):
      return null
      
    // Block to receive the remaining updates
    queriedUpdates = queriedUpdatesFuture
    lastUpdates = [receivedUpdate, previousSelfUpdate]
    lastUpdates.appendAll(queriedUpdates)
    
    // Obtain highest timestamp from the updates
    timestamp = receivedUpdate.getTimestamp()
    for update in lastUpdates:
	if update.getLastUpdate() > timestamp :
	  timestamp = update.getLastUpdate()
	  
    streamCode = currentStream.getCode()
    newUpdate = executeCode(streamCode, lastUpdates, timestamp)
    
    return newUpdates
\end{Code}
\vspace{2ex}

More formally, let $S$ be a composite stream that takes as inputs the SUs
generated by $N$ streams. Let $su^{t_{i}}_{i}$ be the
the most recent SU associated to the $i^{th}$ stream that is a data source for
$S$, where $0 \leq i < N$, and let be $t_{i}$ the associated timestamp to
$su^{t_{i}}_{i}$. Also, let $su^{t_{s}}_{s}$ be the most recent SU
associated to the stream $S$. Notice that it is possible that $\exists_i$ such
that $i = s$ if $S$ consumes its own previously generated data to produce new
outputs.

Then we can define $SU^{t}_{s,in}=\{su^{t_{0}}_{0}, su^{t_{1}}_{1}, \ldots
,su^{t_{n-1}}_{n-1}\}$ as the set of $N$ inputs that $S$ will use to produce one
new output $SU^{t}_{s,out}$ with timestamp $t$. This output will be defined as a
function $SU^{t}_{s,out}=f(SU^{t}_{s,in})$ that is user-defined.

Given these definitions, ServIoTicy needs to guarantee that the function $f$ is 
calculated (and an output $SU^{t}_{s,out}$ emitted) only once for the same set
set of input values, and that at least one of the SUs in $SU^{t}_{s,in}$ needs
to be updated (with a more recent timestamp) to trigger the computation again. 
Furthermore, it is necessary that the set $SU^{t}_{s,in}$ satisfies that
$\exists su^{t_{i}}_{i} \in SU^{t}_{s,in}$ such that ${t_{i}} > t$ to initiate
the computation of $f$ to emit $SU^{t}_{s,out}$.

This restriction can be enforced by checking all the elements of
$SU^{t}_{s,in}$ everytime that an element of the set is updated. But this
approach can result in performing large amounts of costful operations just to
decide that the conditions were not satisfied and that no new output needs to be
emitted.

To mitigate this problem, ServIoTicy relaxes the previously stated restriction
to the form ${t_{j}} > t$ where $0 \leq j < N$ and  $su^{t_{j}}_{j}$ is the
actual element in $SU^{t}_{s,in}$ that triggered the computation. This
relaxation is possible because if  an element exist in the set other than
the one triggering the computation that has a more recent timestamp than $t$,
then this it is very unlikely that this element has been computed before in
time, because then $t$ would have to be as recent as its timestamp. 
Otherwise, if the element with more recent timestamp has not yet triggered the
computation, then it means that the SU has been stored for the source stream and
it must be awaiting in a queue its time to be processed, and therefore it will
trigger the computation soon.

Listing~\ref{list:consistency} summarizes this time-consistency keeping algorithm.

\subsection{Execution trees of the Data Processing Pipelines}
The structure of a pipeline created using the ServIoTicy subscription model
is by definition a directed graph. In practice, though, it behaves more like a
set of trees. The reasoning behind this statement is discussed in this section.

\begin{figure}
  \centering
      \subfigure[]{%
	  \label{fig:topology_async_discard}
		  \includegraphics[width=.20\textwidth]{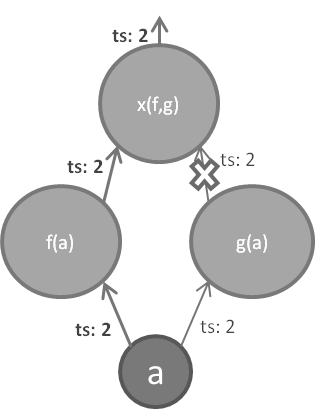}
		  \hspace{1em}
      }%
      \subfigure[]{%
	  \label{fig:topology_cycle_discard}
		  \includegraphics[width=.20\textwidth]{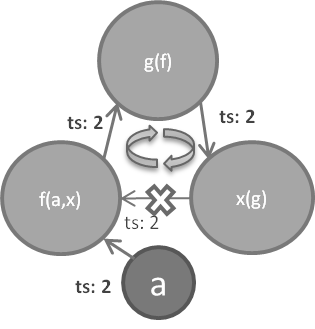}
      }
  \caption{%
      Old data discard}%
   \label{fig:discards}
\end{figure}

When an update reaches a stream, if it is newer than the last generated
update, the computation will be triggered. But if the received update is as new
as the last generated update, the computation will be discarded.
Consider a stream that has several inputs and they originally come from the
exact same entry stream to the pipeline (source). When one of the inputs
receives an update, at some point all the other inputs will receive an update
with the same timestamp and the subsequent computations will always be
discarded. Only the first update to reach the stream will trigger the
computation. An example of this situation can be seen in the
Figure~\ref{fig:topology_async_discard}.

Suppose that all the streams on this pipeline have a SU with timestamp 1 
in their historic data. \emph{a} is the only source of the pipeline, which has 
two streams subscribed, \emph{f(a)} and \emph{g(a)}. Both of them send their 
results to \emph{x(f,g)}, but the SU from \emph{f(a)} is the first one to reach 
\emph{x(f,g)}. The one coming from \emph{g(a)} is discarded because by the time it reaches
\emph{x(f,g)}, there already is in the stored data a SU with timestamp 2 that was
generated using the SUs from \emph{f(a)} with timestamp 2 and from \emph{g(a)} with 
timestamp 1.

This situation is equally valid for cycles, shown in
Figure~\ref{fig:topology_cycle_discard}, as an input closing a cycle shares
exactly the same sources as all the other inputs in the stream.

%

\begin{figure}
  \vspace{-2ex}
  \centering
      \subfigure[Pipeline digraph]{%
	  \label{fig:digraph}
		  \includegraphics[width=.19\textwidth]{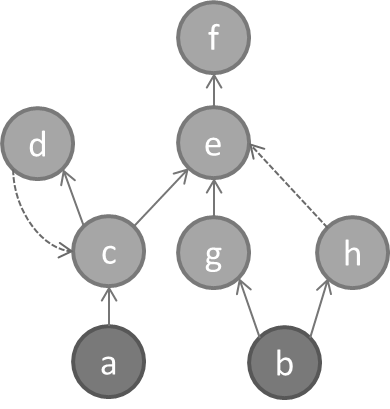}
      }%
      \subfigure[Execution trees]{%
	  \label{fig:trees}
		  \includegraphics[width=.25\textwidth]{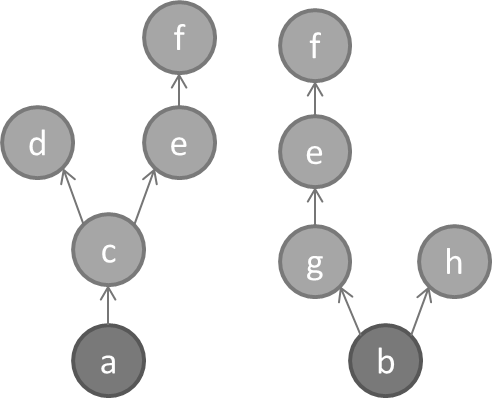}
      }
  \caption{%
      Relation between a pipeline and its execution trees}%
   \label{fig:graphs}
   \vspace{-2ex}
\end{figure}

From this reasoning it can be deduced that the set of paths
of the triggered computations from a single source will 
always end up looking like a tree. For example Figure~\ref{fig:digraph} represents
the graph of a valid pipeline.
The computations that would be generated from the subscriptions 
\emph{d}$\rightarrow$\emph{c} and \emph{h}$\rightarrow$\emph{e} 
are discarded for the explained reasons. Therefore the execution
graphs look like in Figure~\ref{fig:trees}, and updates from \emph{d}
to \emph{c} and from \emph{h} to \emph{e} will only be queried.

Another interesting property of a pipeline is the novelty of its generated
data, and it is useful for evaluating the quality of a stream.
A stream generates novel data when it has an input with a source that no other 
input of the same stream has. The further a stream is in a path from the last 
new source addition, the less novel its generated SUs are. For example in 
Figure~\ref{fig:digraph}, \emph{c}, \emph{g}, \emph{h} and \emph{e} are 1 level more novel
than \emph{f} and \emph{d}. See that \emph{e} gets data sourced on \emph{b}
from two inputs, but theres also another input sourced on \emph{a}. On the 
other hand \emph{f} and \emph{d} are one vertex away from the most novel source.
At the levels of data novelty of this example, getting data from \emph{f}
or \emph{d} is not a problem. The problem comes when the distance from the most novel
stream is too far away
will always take too much time to process an SU that will not add
much value to what it is already evaluated, and will generate several
discarded computations which will end up being time consumed without a result.
Novel data means faster dispatch, less noise in the pipeline and more added
value on the data.

%
%
%
%

\subsection{Runtime implementation and user-code injection}
The software that dispatches the incoming SUs and executes the pipelines runs on
STORM. STORM topologies are static, but the pipelines can easily change over
time, add connections between them, and have arbitrary sizes. For this reason
the STORM topology in ServIoTicy runs the stages described in
Section~\ref{sec:stages}, common to all the pipelines to be processed. On the
subscribers dispatch stage, the target streams are requested, with the code to
be executed in them (previously deployed by the owner of the Service Object
using the REST API).
In the different execution stages (filters and transformation), the JavasScript
code related to it is executed on a JavaScript engine. The JavaScript engine
used is Rhino.
\section{Evaluation}
\label{sec:experiments}

\begin{figure*}[ht!]
     \vspace{-2ex}
     \begin{center}
       \subfigure[Graph]{%
            \label{fig:graph-24}
		    \includegraphics[width=.3\textwidth]{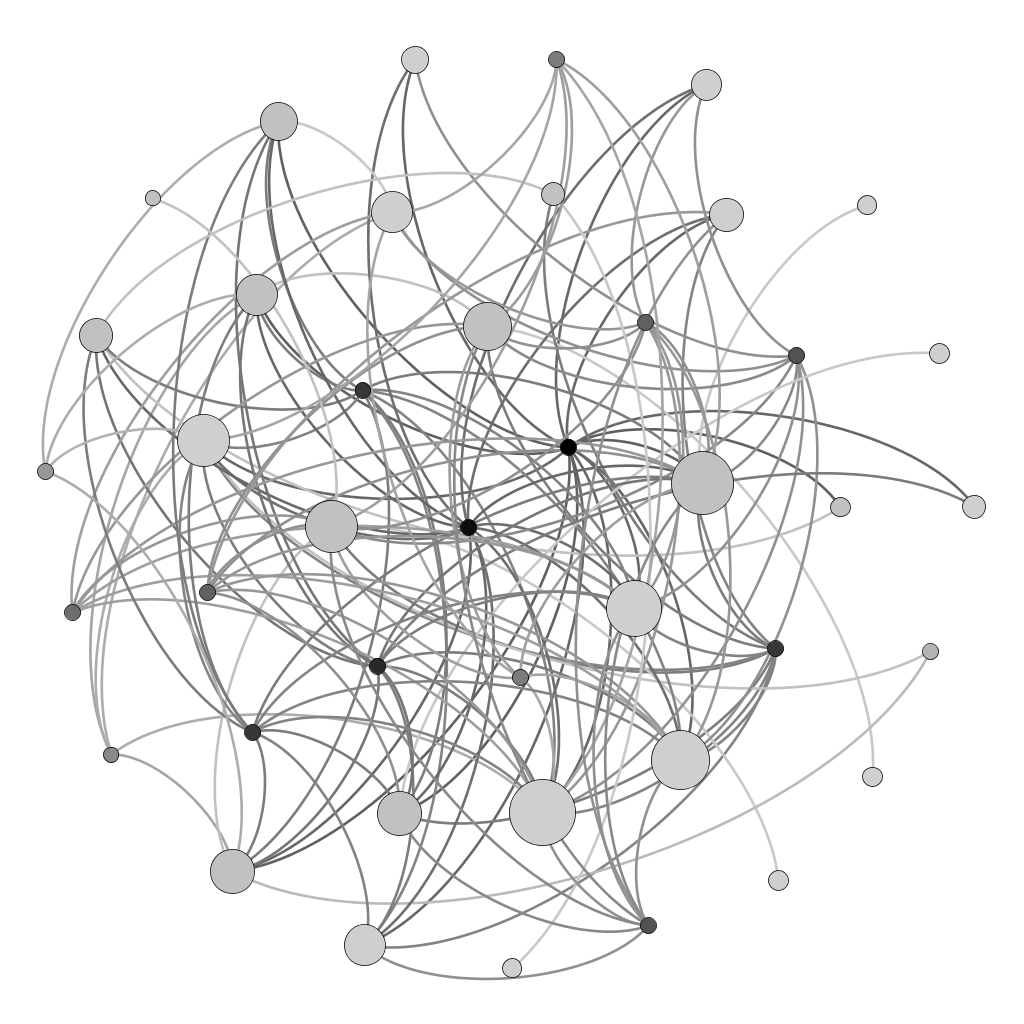}
        }%
       \subfigure[Input stage latencies]{%
            \label{fig:err-24-in}
		    \includegraphics[width=.35\textwidth]{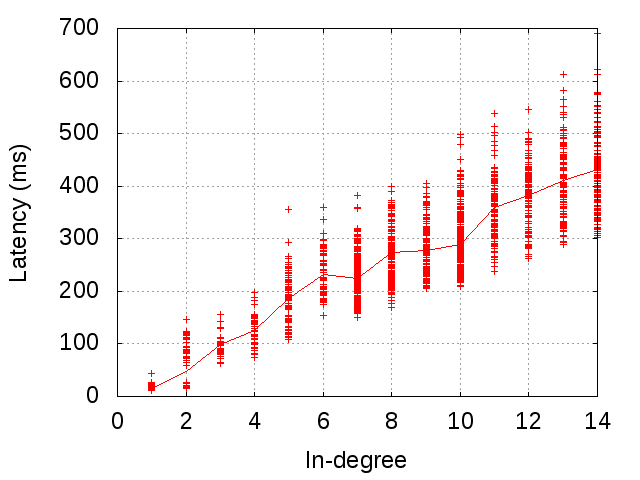}
        }%
        \subfigure[Output stage latencies]{%
            \label{fig:err-24-out}
		    \includegraphics[width=.35\textwidth]{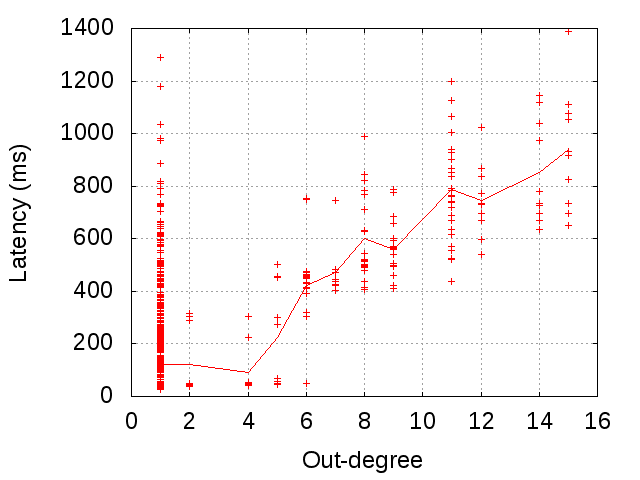}
		    }%
    \end{center}
    \caption{%
        Topology number 3 and its related experiments results}%
   \label{fig:stages-24}
\end{figure*}

\begin{figure}
\centering
\includegraphics[width=0.9\columnwidth]{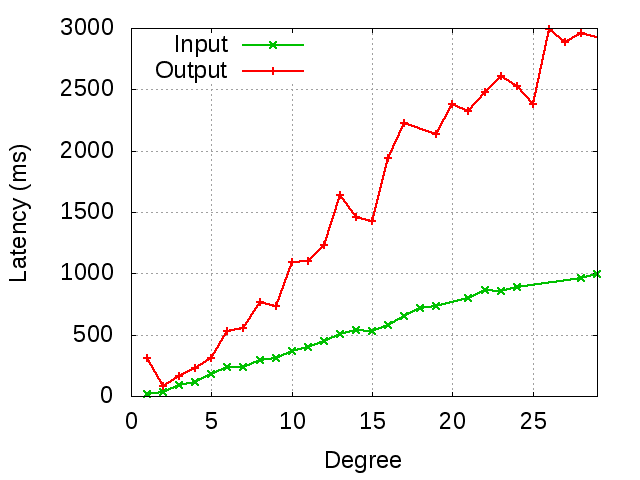}
 \caption{Stage latency by degree}
\label{fig:allstage}
\end{figure}

This section presents a performance evaluation of the
implementation of the ServIoTicy Data Pipelines. 

\subsection{Evaluation Methodology and Infrastructure}

The evaluation is organized in two different experiments. In Experiment
1, we explored the performance of several randomly-generated topologies. We
present here the average results for all of them and the specific results
of one illustrative case. In Experiment 2 individual properties of the graphs, like
depth of the in and out degree for a DPP, were isolated and studied in more
detail. For each experiment, a number of SUs were submitted to the topologies,
and we measured the time it took for each SU to be propagated to all the streams that
were subscribed directly or indirectly to the SU.

To drive the evaluation we developed a tool to automate the generation and
deployment of randomly generated Data Processing Pipelines. The tool provides
several control knobs to customize the properties of the topologies being
generated. The most relevant controls are the number of streams, the number of
composite streams, the number of operands per stream and how the operands are
distributed between the streams.

The tests were run on two sets of nodes: one set for running the client 
emulators and one set for running the servers of the system under test. The 
'server' set was composed of 16 two-way 4-core Xeon L5630  @2.13GHz Linux boxes, 
for a total of 8 cores per node and 16 hardware threads because hyperthreading 
was enabled. Each 'server' machine was enabled with 24GB of RAM. The 'client' 
set was composed of 2 two-way 6-core Xeon E5-2620 0 @2.00GHz Linux boxes, for a 
total of 12 cores per node and 24 hardware threads because hyperthreading was 
enabled. Each 'server' machine was enabled with 64GB of RAM. All nodes were 
connected using GbE links to a non blocking 48port Cisco 3750-X switch.
The ServIoTicy data processing runtime was deployed on 2 server
machines, and 1 client machine was used to generate the SUs. The REST API used
the other nodes to host its components.
For the data processing
pipelines we used Apache STORM v0.9.2-incubating, Kestrel v2.4.1 and ZooKeeper
v3.4.5. 


\subsection{Experiment 1}
For this experiment, we generated six different testing topologies for
ingesting data produced by a Service Object. The characteristics of these
topologies are summarized in Table~\ref{tab:random-topologies}. They can be
grouped based on their size (small, medium or large), and we randomly produced 2
samples of each complexity level. Based on our experience, topologies 1 and 2
emulate two realistically sized situations. Topologies 3 and 4 are large cases.
Finally, topologies 5 and 6 are extreme cases. 
A graphical representation of topology number 3 is shown  in
Figure~\ref{fig:graph-24}. In this figure, dark nodes indicate a high out-degree and big nodes represent high in-degree.
The in and out degree related properties are also very relevant
for this experiment, as they have a big impact on the metrics taken.

\begin{center}
    \begin{table}
    \begin{center}
    \caption{Pseudo-random topologies} \label{tab:random-topologies}
    \begin{tabular}{ | l | l | l || l | l || l | l | p{5cm} |}
    \hline
    Type & \multicolumn{2}{|c|}{Small} & \multicolumn{2}{|c|}{Medium} &
      \multicolumn{2}{|c|}{Big}\\
    \hline
    Id & 1 & 2 & 3 & 4 & 5 & 6\\
    \hline
    \hline
    Max in-degree & 9 & 8 & 14 & 16 & 29 & 24\\
    Mean in-degree & 1.42 & 1.94 & 3.54 & 3.51 & 5.28 & 6.18\\
    In-degree std. dev. & 2.22 & 2.63 & 4.36 & 5.05 & 7.43 & 7.38\\
    \hline
    Max out-degree & 4 & 7 & 15 & 15 & 25 & 28\\
    Mean out-degree & 1.42 & 1.94 & 3.54 & 3.51 & 5.28 & 6.18\\
    Out-degree std. dev. & 1.07 & 2.14 & 4.59 & 4.44 & 7.71 & 9.48\\
    \hline
    Edges & 30 & 37 & 149 & 151 & 423 & 458\\
    Nodes & 21 & 19 & 42 & 43 & 80 & 74\\
    \hline
    Sources & 11 & 9 & 17 & 18 & 30 & 24\\
    Sinks & 4 & 7 & 15 & 15 & 25 & 28\\
    \hline
    Density & 0.14 & 0.21 & 0.17 & 0.16 & 0.13 & 0.16\\
    Connectivity& 1 & 1 & 1 & 1 & 1 & 1\\
    Edge-connectivity & 1 & 1 & 1 & 1 & 1 & 1\\
    \hline
    \end{tabular}
    \end{center}
    \end{table}
\end{center}

For each data source, 10 Sensor Updates were sent to the platform in sequence: a new update
was generated only after the previous pipeline computation was finished. During the topology
execution, two metrics were measured for each stream. The first metric is the
execution time to perform all the data queries required to complete the
processing, named the {\it input stage}.
This metric measures the effect of using several inputs to generate a new
update.
The second metric is the time difference between the instant at which a new update is emitted and the time at which 
all subscribers have received it: this metric measures how the topology processing time is affected 
by the number of subscribers at each stage of the processing pipeline. This is
named in this section as the {\it output stage}.

Other stages were also measured, such as the injected code processing time or the time an
update remained in the Kestrel queue. The function to generate a new
update was always a summation of the inputs, and so had complexity $O(n)$, 
being $n$ the in-degree. However, these measures resulted on negligible times and have not been included in the discussion.

Figures~\ref{fig:err-24-in} and~\ref{fig:err-24-out} show all the latencies
measured for topology number 3. Each dot in the plot represents one execution of a
topology node with a given in- or out-degree that corresponds to the value in
the X-axis. The average latency for each degree is also drawn in both charts as
a solid line. As it can be observed, latency grows linearly with the degree
level as some sequential operations are required for each operation. Although
the communication is made asynchronous, the stages need to be closed before
jumping to the next step for the topology, and therefore it is necessary to wait
for all on-the-fly operations to complete at some point, what results in a
waiting time that is proportional to the number of initiated operations and
therefore the degree of the stage.

Finally, Figure~\ref{fig:allstage} shows the average latency on the input and
output stages for every related degree, across all six topologies.
As it can be observed, the latency of both the input and output stages grow
linearly, but in a higher pace in the output stage.
While the in-degree latencies look almost the same to
Figures~\ref{fig:err-24-in}, the out degree grows
faster. The reason for this worse performance is that
this Figure reports average values that are affected by the higher
latencies of the bigger topologies. Therefore, the time of the output stage not only
depends on the out-degree, but also on the total size of the topology. And in
particular, as it will be shown in the next experiment, the topology length is
the most important factor that affects the performance of the topologies. The
larger the topology is, the more operations are run in parallel in the topology
and therefore the largest the response times of the components, resulting in a
slightly higher latency to complete the processing of an update.

\subsection{Experiment 2}
\begin{figure*}[ht!]
   \vspace{-2ex}
     \begin{center}
       \subfigure[Length]{%
            \hspace{-1em}
            \label{fig:pipe-long}
            \raisebox{5mm}{\includegraphics[width=.3\textwidth]{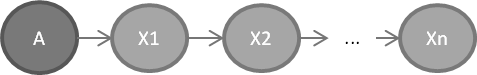}\vspace{-2ex}}
		    \hspace{3em}
        }%
       \subfigure[In-degree]{%
            \label{fig:pipe-in}
		    \includegraphics[width=.2\textwidth]{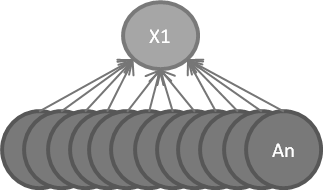}\vspace{-2ex}
		    \hspace{3em}
        }%
        \subfigure[Out-degree]{%
            \label{fig:pipe-out}
		    \includegraphics[width=.2\textwidth]{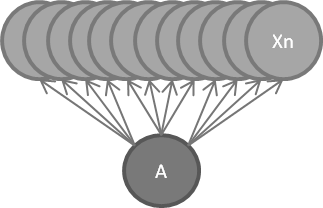}\vspace{-2ex}
		    \hspace{3em}
		    }%
    \end{center}
    \caption{%
        Types of tested pipeline, each one maximizing a property}%
   \label{fig:pipe-experiments}
\end{figure*}
Following the results of Experiment 1, a second experiment was performed to
separately stress the importance of the in degree, the out degree and the
length of a topology path.
The latter measure is also stressed because it can not be parallelized, and so
affects greatly to the overall topology execution.

For the second experiment, three groups of 100 pipelines were deployed, each one 
emphasizing one of the following main properties shown in Figure~\ref{fig:pipe-experiments}:
\begin{itemize}
  \item \emph{Length}: The length of a pipeline is the maximum number of composite streams 
  from one of the sources to any sink. It affects the performance of the pipeline because 
  each one of these streams depends on the result of the previous one, so there is no 
  possible parallelism.
  \item \emph{Out-degree}: A pipeline's out-degree is the average number of subscribers (operators) its 
  streams have. This is directly affected by the parallelism, the less available threads on the machines
  the more it will influence negatively the performance.
  \item \emph{In-degree}: The in-degree is the average number of subscriptions (operands) 
  its streams have. It alters the performance of the execution of a single stream, mainly. The reason is
  that having a big amount of operands in a composition function means that there are more SU queries to
  perform. The impact on the performance of the in-degree will depend on the number of available 
  threads, because the set of queries are asynchronous.
\end{itemize}

Each pipeline in a group exhibits a different number of streams, ranging from 2 to 101. 
In the case of the 'in-degree' type, the pipelines ranged
from 1 source and 1 sink to 100 sources and 1 sink. 'Out-degree' type
ranged from 1 source and 1 sink to 1 source and 100 sinks.
Finally, the 'length' type goes from 1 source and 1 sink only to 1 source and 1
sink with 99 intermediate chained composite streams. This makes 300 pipelines tested.

For each pipeline, 10 Sensor Updates were sent to the platform, at a rate of one
SU per second. During the time all SUs were propagated, several metrics were
collected on the ServIoTicy runtime to determine the delays introduced at each
stage and the end-to-end time to process every SU generated.
\begin{figure}
\centering
\includegraphics[width=1\columnwidth]{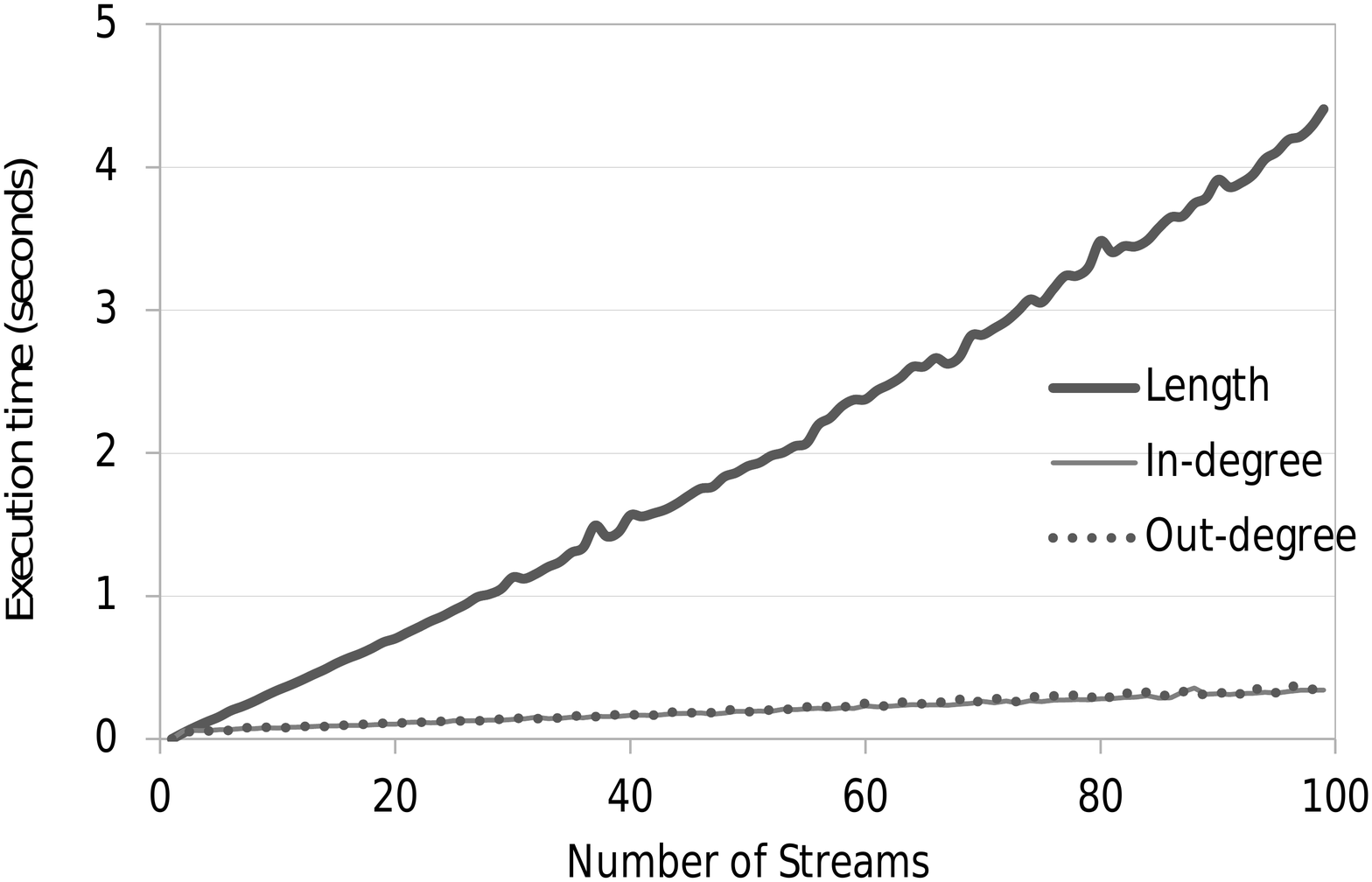}
 \caption{Time to dispatch a SU through an entire topology.}
\label{fig:topology_performance}
\vspace{-3ex}
\end{figure}

Figure~\ref{fig:topology_performance} shows the average total execution time of all the
pipelines, for each one of the 3 types of pipelines considered.
As it can be observed, for the three cases the execution time grows linearly with
the number of streams. 

As it was expected, the time to propagate a SU through
the entire pipeline grows significantly more for the 'length' pipelines because
they can not take advantage of any parallelism: all streams are calculated
sequentially because they contain sequential data dependencies that can not be
skipped.

For the in-degree and out-degree pipelines, it can be observed that the there is 
almost no difference on how the execution time is affected. In comparison with 
Figure~\ref{fig:stages-24}, the latencies are 
significantly lower. Specially in the case of the out-degree, that in this 
last experiment had a mean latency of 350ms to complete a pipeline with out-degree
100. Notice that in Figure~\ref{fig:stages-24}, the mean latency value for
the output stage with out-degree 15 is 950ms. Yet the bigger topology on
this experiment is not far from the one of Figure~\ref{fig:stages-24} in terms of 
subscriptions, and is bigger in terms of streams. The output stage of a stream is then 
highly affected by the longitude of the pipeline, and not by the overall size as concluded
on the first experiment.  Nodes with distance 1 from the
source will end up competing for resources with nodes with distance higher than 1
if the initial out-degree is high enough. There is room for improvement by
prioritizing nodes near to the sources, otherwise some paths on the pipeline
will be faster than others.




\section{Related Work}
\label{sec:related}
In the last years several stream processing platforms have emerged, being Storm~\cite{storm}
the most popular and it is used in this contribution as a platform runtime. 
Storm is a distributed, reliable, and
fault-tolerant stream processing system, which was open sourced by Twitter after acquiring
BackType and now distributed by the Apache Software foundation. ZeroMQ or Netty are the
messaging interfaces between the computation units. In the last versions 
multi-tenancy was added in terms of several tenants deploying isolated topologies.
This topologies are always in memory whether are being used or not, and there is
not data subscription between tenants. Also open-source and distributed by the 
Apache Software Foundation are Apache Samza~\cite{samza} and Apache Flink~\cite{flink} and Apache S4~\cite{s4}.
Apache Samza uses Kafka for the whole messaging between the computation 
units and YARN for resource management. Apache Flink is a streaming dataflow 
engine that provides data distribution, communication, and fault tolerance 
for distributed computations over data streams. It has two APIs, one for
data streams and another for data sets or batch processing. Flink also bundles 
libraries for domain-specific use cases like complex event processing and machine
learning. Apache S4 is an already deprecated project started by Yahoo with a very
similar topology based philosophy to Storm and an architecture resembling the Actors model.
Microsoft Research developed a proprietary solution for complex event processing
called StreamInsight~\cite{streaminsight}. It also leverages a programing model
for temporal data streams, operator algebra and continuous queries. Other relevant
foundations on stream processing in real-time from Microsoft come the 
CEDR~\cite{barga2006consistent} project. It is centered in the problem of 
keeping time consistency on event streaming.
Other well known research related projects on data streams are Aurora~\cite{aurora} and 
its forks Medusa~\cite{medusa} and Borealis~\cite{borealis}. None of this projects are
maintained anymore.
From the perspective of data stream sharing, StreamGlobe~\cite{kuntschke2005streamglobe}
offers a Grid Computing solution using a P2P approach. It consist then in stream sharing
between machines but not multi-tenancy.

Data Centric view of the IoT is not something new for ServIoTicy as it was
widely covered in the survey presented in~\cite{article:thingsmatter}. What
ServIoTicy uniquely provides is an open source solution that challenges the
features of commercial solutions such as Xively~\cite{xively} and
Evrythng~\cite{evrythng}, while extending their capabilities with the ability to
inject user-defined code into its stream processing runtime. 

There are other
open source platforms for IoT in the market, but they are focused on other
aspects of the Internet of Things. The DeviceHive~\cite{devicehive} project
offers a machine-to-machine (M2M) communication framework for connecting devices
to the IoT. It includes easy-to-use Web-based management software
for creating networks, applying security rules and monitoring devices.
Devicehub.net~\cite{devicehub} is a cloud-based service that stores IoT-related
data, provides visualizations of that data and allows users to control IoT
devices from a Web page. The IoT Toolkit~\cite{iottoolkit} project provides a
variety of tools for integrating multiple IoT-related sensor networks and
protocols. The primary project is a Smart Object API, but it also aims to
develop an HTTP-to-CoAP Semantic mapping.
Mango~\cite{devicehub} is a popular open source Machine-to-Machine (M2M)
software, which is web-based and supports multiple platforms. Key features
include support for multiple protocols and databases, and user-defined events
among others. Nimbits~\cite{nimbits} can store and process a specific type of
data previously time- or geo-stamped. A public platform as a service is
available, but it can also be downloaded and deployed on Google App Engine, any
J2EE server on Amazon EC2 or on a Raspberry Pi.
Netquest~\cite{netquest} is a programming model to ease the development
of ubiquitous applicactions on sensor networks.
On paper~\cite{imote}, Netquest is used to work on a small network of iMote devices.
OpenRemote~\cite{operemote} offers four different integration tools for
home-based hobbyists, integrators, distributors, and manufacturers. It supports
dozens of different existing protocols, allowing users to create nearly any kind
of smart device they can imagine and control it using any device that supports
Java. The SiteWhere~\cite{sitewhere} project provides a complete platform for
managing IoT devices, gathering data and integrating that data with external
systems. SiteWhere releases can be downloaded or used on Amazon's cloud. It also
integrates with multiple big data tools, including MongoDB and Apache HBase.
Finally, ThingSpeak~\cite{thingspeak} can process HTTP requests and store and
process data. Key features of the open data platform include an open API,
real-time data collection, geolocation data, data processing and visualizations,
device status messages and plugins.

Deployment of IoT platforms on the Cloud is also covered in the literature.
In~\cite{iotcloud}, authors propose strategies for deciding the best approach at
the time of making cloud-based deployments of IoT applications using nowadays
regular cloud technologies. Another recent work~\cite{cloudpubsub} studies the
implementation of IoT platforms on top of cloud-based pub/sub communication
infrastructures. Finally, authors go one step beyond in~\cite{iotsde} by
leveraging completely Software Defined Environments for managing the Cloud
infrastructures in which IoT applications are deployed.

\section{Conclusions and Future Work}
\label{sec:futurework}
In this paper we have introduced 
a multi-tenant data stream processing mechanism on top of Apache STORM that 
enables the tenants to share data streams between them.
STORM provides auto-scaling capabilities that make it particularly suitable for
cloud deployments.
The ServIoTicy runtime allows for users to deploy custom service codes
inside Service Objects in the form of composite streams, and subscribe those
streams to multiple sources of data (either outside the platform on real-world
devices or in other streams defined in the ServIoTicy platform by other users).
The user-code will be automatically injected in the STORM topology and
executed when a unit of data is generated from a source to which the composite
stream is subscribed. 
The runtime is designed to be highly scalable, following a lock-free model that
combines operations triggered by new data being generated inside or outside the
platform, with queries performed over historic data logged for existing Service
Objects. The design imposes some restrictions mainly related to the timestamps
of the updates being processed, and some optimizations are applied to improve
the scalability of the platform.
A basic evaluation of the runtime is included in this work, showing how
acceptable response times of less that 100ms can be delivered by basic composite
streams, and that for most realistic pipelines can be processed in the range of
less than a second.
The work presented in this paper is, to our knowledge, the first multi-tenant
IoT data processing platform for the Cloud.

The next steps to follow after this contribution will be to extend the programing
model to enable some new features. 
One of them is having sliding window aggregators defined by static size, time interval and random
events. Being this the scenario of data streams in real-time, the programing model
needs to enforce efficient incremental algorithms for the aggregators so
the computation time with millions of updates is ideally lower than the interval between
the arrival of each update. 
Moreover, another interesting feature is dynamic data stream subscriptions. To subscribe
to one or several streams it is needed to provide their unique ids. A more flexible way
to do that would be subscribing dynamically to the streams that match some specific
features. Every time a stream is added to the platform and it matches a dynamic 
subscription criteria, it will be binded automatically to its subscribers.
\section*{Acknowledgments}
This work is supported by the European Research Council (ERC) under the European Union's Horizon 2020 research and innovation programme (grant agreement no. 639595); de Catalunya under contract 2014SGR1051; the ICREA Academia program; and the BSC-CNS Severo Ochoa program (SEV-2015-0493); the Spanish Ministry of Economy under contract TIN2015-65316-P and the Generalitat.


%

%
%

\ifCLASSOPTIONcaptionsoff
  \newpage
\fi



%

\bibliographystyle{IEEEtran}
\bibliography{IEEEabrv,references}

\end{document}